\documentclass{article}
\usepackage{conf,amsmath,epsfig}

\newcommand{\be}{\begin{equation}}  
\newcommand{\ee}{\end{equation}}  

\def\x{{\mathbf x}}
\def\w{{\mathbf w}}

\title{NEURAL CRYPTOGRAPHY}

\twoauthors
  {Wolfgang Kinzel}
   {Institute for Theoretical Physics\\
   University, Am Hubland\\
   97074 W\"urzburg, Germany}
  {Ido Kanter}
   {Minerva Center and Department of Physics\\
   Bar-Ilan University\\
   52100 Ramat-Gan, Israel}

\begin{document}
%
\maketitle
\begin{abstract}

Two neural networks which are trained on their
  mutual output bits show a novel phenomenon: The networks synchronize to a
  state with identical time dependent weights. It is shown how
  synchronization by mutual learning can be applied to cryptography: secret key
  exchange over a public channel.
\end{abstract}
\section{Introduction}
\label{sec:intro}

Neural networks learn from examples. This concept has extensively been
investigated using models and methods of statistical mechanics
\cite{HeKrPa,EnVa}. A ''teacher'' network is presenting input/output
pairs of high dimensional data, and a ''student'' network is being
trained on these data. Training means, that synaptic weights adopt by
simple rules to the input/output pairs. After the training phase the
student is able to generalize: It can classify -- with some
probability -- an input pattern which did not belong to the training
set.

Training is a dynamic process. The examples are generated step by step
by a static network - the teacher. The student tries to move towards
the teacher. It turns out, that for a large class of models the
dynamics of learning and generalization can be described by ordinary
differential equations for a few order parameters \cite{BiCa}.

Recently this scenario has been extended to the case of a dynamic
teacher: Both of the communicating networks receive an identical input
vector, generate an output bit and are trained on the corresponding
bit of their partner. The analytic solution shows a novel phenomenon:
synchronization by mutual learning \cite{MeKiKa}. The synaptic weights
of the two networks relax to a common identical weight vector which
still depends on time.  The biological consequences of this phenomenon
are not explored, yet, but an interesting application in cryptography
has been found: secure generation of a secret key over a public
channel \cite{KaKiKa}.

In the field of cryptography, one is interested in methods to transmit
secret messages between two partners A and B. An opponent E who is able
to listen to the communication should not be able to recover the secret
message.

Before 1976, all cryptographic methods had to rely on secret keys for
encryption which were transmitted between A and B over a secret channel
not accessible to any opponent. Such a common secret key can be used,
for example, as a seed for a random bit generator by which the bit
sequence of the message is added (modulo 2).

In 1976, however,  Diffie and Hellmann found that a common secret key
could be created over a public channel accessible to any opponent. This
method is based on number theory: Given limited computer power, it is
not possible to calculate the discrete logarithm of sufficiently large
numbers \cite{St}.

Here we show how neural networks can produce a common secret key by
exchanging bits over a public channel and by learning from each other
\cite{KaKiKa,RoKaKi,Ro}.

\section{Training the tree parity machine}

\begin{figure}[htb]
  \centering
 \centerline{\epsfig{figure=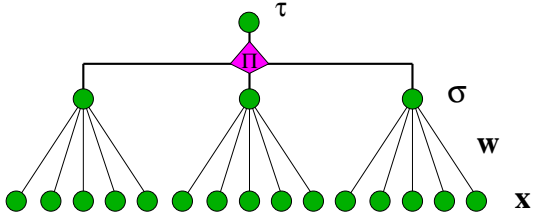,width=7cm}}
\caption{Parity machine with three hidden units.}
\label{par}
\end{figure}

Both of the communicating partners A and B are using a multilayer
network with K hidden units: A tree parity machine, as shown in figure
\ref{par}. In this paper we use K=3, only. Each network consists of
three units (perceptrons, i=1,2,3):
\[
\sigma_i^{A} = \mbox{sign} (\w_i^A \cdot \x_i); \quad \hfill
\sigma_i^{B} = \mbox{sign} (\w_i^B \cdot \x_i)
\label{eins}
\]
The $\w$ are N-dimensional vectors of synaptic weights and the $\x$
are N-dimensional input vectors.  Here we discuss discrete weights and
inputs, only:
\[
w_{i,j}^{A/B}  \in \{ -L, -L +1, ... , L-1, L \}; \quad x_{i,j} \in \{-1,+1\}  
\]
The three hidden bits $\sigma$ are combined to an output bit
$\tau$ of each network:
\[
\tau^{A}  =  \sigma^{A}_{1} \; \sigma^{A}_{2} \; \sigma^{A}_{3};  \quad
\tau^B  =  \sigma^{B}_{1} \; \sigma^{B}_{2} \; \sigma^{B}_{3}
\]

The two output bits $\tau$ are used for the mutual training process.
At each training step the two machines A and B receive identical
input vectors $\x_1 , \x_2 , \x_3$. The
training algorithm is the following: Only if the two output bits are
identical, $\tau^{A} = \tau^B$,  the weights can be changed. In this
case, only the hidden unit $\sigma_{i}$ which is identical to $\tau$
changes its weights using the Hebbian rule
\[
\w^{A}_{i} (t+1) = \w^{A}_{i} (t) + \x_i 
\]
and the same for the network B.
If  this training step pushes any component $w_{i,j}$  out of the interval 
${-L,...,L}$ the component is replaced by $\pm L$, correspondingly.

Consider for example the case $\tau^{A} = \tau^B = 1$. There are four possible
configurations of the hidden units in each network:\\ 
$(+1, +1, +1), (+1,-1, -1), (-1, +1, -1), (-1, -1, +1)$\\
In the first case, all three weight vectors $\w_1,
\w_2, \w_3$ are changed, in all other three
cases only one weight vector is changed. The partner as well as any
opponent does not know which one of the weight vectors is updated.

Note that  the two multilayer networks may be considered as a
system of random walks with reflecting boundaries. Each of the $6 N$
components $w_{i,j}$ of the weight vectors moves on $2 L +1$ lattice points.
$w_{i,j}$ makes a step $x_{i,j}=\pm 1$ if the corresponding global signals
$\tau$ and $\sigma$ allow this. If it hits a boundary it is reflected. 
Since any two weights $w^A_{i,j}$ and  $w^B_{i,j}$ receive an identical
input $x_{i,j}$, every common step where one component is reflected
decreases the distance between the two weights. 
As we will see in the following section, this finally results in
identical weight vectors. 
 
\section{GENERATION OF SECRET KEYS}

Mutual learning of tree parity machines, as explained before, leads to
synchronization of the time dependent synaptic vectors $\w^A_i$ and
$\w^B_i$. This is the result of numerical simulations as well as
analytic solutions of the model\cite{KaKiKa,RoKaKi,Ro}. Both partners start
with random weight vectors (3 N random numbers each) and train their
weight vectors according to the algorithm explained above. At each
training step they receive three common random input vectors $\x_i$.

It turns out that after a relatively short number of training steps
all pairs of the weight vectors are identical, $\w^A_i = \w^B_i$.  The
two multilayer networks have identical synaptic weights.  Since,
according to the learning rule, after synchronization at least one
pair of weight vectors is changed for each training step, the synaptic
weights are always moving. In fact, it is hard to distinguish this
motion form a random walk in weight space\cite{Metzler}. Therefore the
two multilayer networks perform a kind of synchronized random walk in
the discrete space of $(2L+1)^{3 N}$ points.

\begin{figure}[htb]
\centering
\includegraphics[width=7cm]{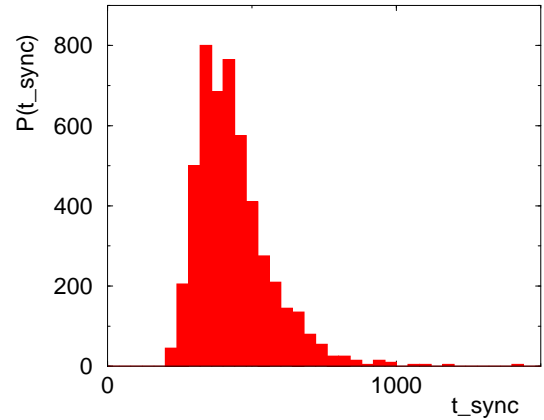}
\caption[]{Distribution of synchronization time for $N=100, L=3$.}
\label{tsync}
\end{figure}

Figure \ref{tsync} shows the distribution of synchronization time for
$N=100$ and $L=3$. It is peaked around $t_{sync} \simeq 400$.  After
400 training steps each of the 300 components of the network A has
locked into its  identical counterpart of the network B. One finds that the
average synchronization time is almost independent on the size $N$ of
the networks, at least up to $N=10000$. Asymptotically one expects 
an increase like $\log N$. 

Synchronization of neural networks can immediately be translated to
key generation in cryptography: The common identical weights of the
two partners A and B can be used as a key for encryption, either
immediately as one-time pad, as a seed for random bit generators
or as a key in other encryption algorithms (DES,AES)\cite{St}. 

Compared to algorithms based on number theory, the neural algorithm
has several advantages: First, it is very simple. The training
algorithm is essentially a linear filter which can easily implemented
in hardware.  Second, the number of calculations to generate the key
is low. To generate a key of length $N$ one needs of the order of $N$
computational steps. Third, for every communication, or even for
every block of the message, a new key can be generated. No secret
information has to be stored for a longer time.

But useful keys have to be secure. An attacker E who is recording the
communication between A and B should not be able to calculate the secret key.
Attacker will be discussed in the following.

\section{ATTACKS}

A secure key exchange protocol should have the following property: Any
attacker who knows all of the details of the protocol and all of the
information exchanged between A and B should not have the
computational power to calculate the secret key.

We assume that the attacker E knows the algorithm, the sequence of
input vectors and the sequence of output bits. In principle, E could
start from all of the $(2L+1)^{3N}$ initial weight vectors and
calculate the ones which are consistent with the input/output
sequence. It has been shown, that all of these initial states move
towards the same final weight vector, the key is unique \cite{Ur}.
However, this task is computationally infeasible.

Hence one has to find an algorithm which tries to adapt to the known
input/output. Note that the training rule for A and B has the
property: If a pair of units is synchron it remains so forever. The
synchronous state is an attractor of the learning dynamics.
Any algorithm for the attacker E should have this property, too.

An immediate guess for a possible attack is the following: E uses the
same algorithm as one of the partners, say B. If $\tau^A=\tau^B$ the
weight vectors of E are changed for which the unit $\sigma^E_i$ is
identical to $\tau^A$.

\begin{figure}[htb]
\centering
\includegraphics[width=7cm]{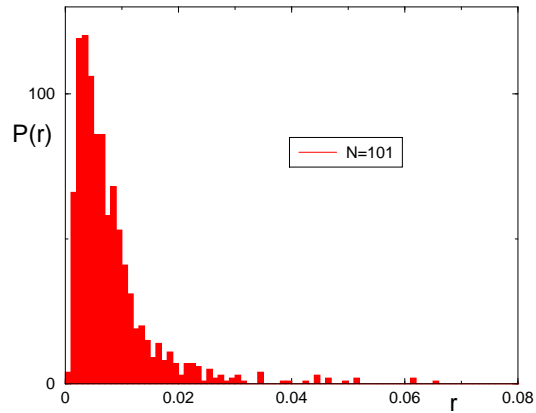}
\caption[]{
  Distribution of the ratio of synchronization time between networks A
  and B to the learning time of an attacker E.}
\label{rat2}
\end{figure}

In fact, numerical simulations as well as analytic calculations show
that an attacker E will synchronize with A and B after some learning
time $t_{learn}$\cite{KaKiKa,RoKaKi,Ro}. However, the learning time is much
longer than the synchronization time. Figure \ref{rat2} shows the
distribution of the ratio between synchronization and learning times.
On average, learning is about 1000 times slower than synchronization.
But even the tail of the distribution never exceeded the factor 10
(for 1000 runs).  Therefore, if the training process is stopped
shortly after synchronization, the attacker has no chance to calculate
the key.  The key is secure for this algorithm of attack.

Why does this work at all? What is the difference between the partner
B and the attacker E, who both have the same information? The reason
is that B can influence the network A whereas E can only listen.
Synchronization as well as learning is a competition of attraction and
repulsion controlled by the output bits. One can show, for the parity
machine the probability for repulsion is much larger for E than for A
and B, at least close to synchronization. This is not true for the
committee nor the simple perceptron\cite{St,Ro}.

However, one cannot exclude that E finds attacks which perform better
than the simple attack described above. In fact, recently several
attacks were found which seem to crack the key exchange\cite{Sh}. The most
successful one has two additional ingredients: First, an ensemble of
attackers is used. Second, E makes additional training steps when A
and B are quiet, $\tau^A \ne \tau^B$.

An ensemble is helpful if the distribution of learning times is broad.
Then there may be a chance that some of, say 10000, attackers will
synchronize before A and B. If one reads all of the 10000 encrypted
messages one will detect the key from those messages which have a
meaning.  The additional training step goes as follows: If $\tau^E \ne
\tau^A$ search for the unit with smallest internal field $\w^E_i \cdot
\x_i$, flip the corresponding $\sigma^E_i$ and proceed with training
as above. This step enforces learning by changing only the information
which is close to the decision boundary.

This algorithm succeeds to find the key for the value $L=3$.  There is
a nonzero fraction $P(L)$ of attackers which synchronize with the two
partners A and B \cite{Sh}. However, a detailed numerical calculation of the
scaling of key generation showed that this fraction $P(L)$ decreases
exponentially fast with the number $L$ of weight values\cite{L}. The
synchronization time, on the other hand, increases only like $L^2$, as
expected form the random walk analogy. Therefore, in the limit of
sufficiently large values of $L$ neural cryptography is secure.

In addition, it has been shown that key generation by mutual learning
can be made even more secure by combining it with synchronization of
chaotic maps\cite{Chaos}.

\section{SUMMARY}

Interacting neural networks are able to synchronize. Starting from
random initial weights and learning from each other, two multilayer
networks relax to a state with time dependent identical synaptic
weights.

This scenario has been applied to cryptography. Two partners A and B can
generate a secret key over a public channel by training their parity
machines on the output bits of their partner. A and B did not exchange
any information over a secret channel before their communication.
Although an attacker can record the communication and knows the
algorithm she is not able to calculate the secret common key which A and
B use for encryption.

This holds for all attackers studied so far. Of course, one  cannot prove
that no  algorithms exist for a successful attack. Future has to show whether
neural cryptography remains secure for more advanced attacks.

To our knowledge, neural cryptography is the first algorithm for key
generation over public channels which is not based on number theory.
It has several advantages over known protocols: It is fast and simple,
for each messages a new key can be used and no information is stored
permanently. Therefore neural cryptography may lead to novel applications
in the future.


\begin{thebibliography}{99}

\bibitem{HeKrPa} J. Hertz, A. Krogh, and R. G. Palmer:    
\emph{Introduction to the Theory of Neural Computation},     
(Addison Wesley, Redwood City, 1991)    

\bibitem{EnVa} A. Engel, and C. Van den Broeck: \emph{Statistical
    Mechanics of Learning}, (Cambridge University Press, 2001)
 
\bibitem{BiCa} M. Biehl and N. Caticha: Statistical Mechanics of On-line
  Learning and Generalization, \emph{The Handbook of Brain Theory and
    Neural Networks}, ed. by M. A. Arbib (MIT Press, Berlin 2001)
    


\bibitem{MeKiKa} R. Metzler and W. Kinzel and I. Kanter, \emph{Interacting
    neural networks}, Phys. Rev. E \textbf{62}, 2555 (2000)
  
\bibitem{Metzler} R. Metzler, W. Kinzel, L. Ein-Dor and I. Kanter, 
\emph{Generation of antipredictibable time series by
    neural networks}, Phys Rev. E 63, 056126 (2001)

  
\bibitem{KaKiKa} I. Kanter, W. Kinzel and E. Kanter, \emph{Secure exchange
    of information by synchronization of neural networks}, Europhys.
  Lett. \textbf{57}, 141-147 (2002)

\bibitem{St} D. R. Stinson, \emph{Cryptography: Theory and Practice}      
(CRC Press 1995)    


\bibitem{RoKaKi} M. Rosen-Zvi, I. Kanter and W. Kinzel, \emph{Cryptography
    based on neural networks: analytical results}, cond-mat/0202350
  (2002)

\bibitem{Ro} M. Rosen-Zvi, E. Klein, I. Kanter and W. Kinzel, \emph{Mutual
    learning in a tree parity machine and its application to
    cryptography}, Phys. Rev. E (2002)

\bibitem{Ur} R. Urbanczik, private communication
  
\bibitem{Sh} A. Klimov, A. Mityagin and A. Shamir, \emph{Analysis of neural
    cryptography}, to be published
  
\bibitem{L} R. Mislovaty, Y. Perchenok, I. Kanter and W. Kinzel, \emph{A
    secure key exchange protocol}, to be published

\bibitem{Chaos} I. Kanter, unpublished

\end{thebibliography}
\end{document}